\documentclass{osa-article}
\journal{oe}
\articletype{Research Article}

\usepackage{todonotes}
\usepackage{changes}
\usepackage[abbreviations]{siunitx}
\usepackage{mathrsfs}

\definechangesauthor{LL}
\definechangesauthor{JL}
\definechangesauthor{SH}

\newcommand{\I}{\textup{i}}

\newcommand{\D}{\textup{d}}
\newcommand{\mR}{\mathbb{R}}

\newcommand{\MTEXT}[1]{\;\;\;\;\;\text{#1}\;\;\;\;\;}
\newcommand{\MLTEXT}[1]{\text{#1}\;\;\;\;\;}

\newcommand{\FT}{\mathcal{F}}
\newcommand{\Fresnel}{\mathcal{D}}
\newcommand{\LinOp}{\mathcal{L}}

\newcommand{\NLOp}{\mathcal{N}}
\newcommand{\NLOpa}[1]{\NLOp_{#1}}
\newcommand{\LinTikhFunc}{\mathscr{T}_{\textup{Lin}}}
\newcommand{\NLTikhFunc}{\mathscr{T}_{\textup{NL}}}

\newcommand{\norm}[1]{\| #1 \|} 

\newcommand{\ip}[2]{\langle #1, \, #2\rangle} 

\newcommand{\Norm}[1]{\left\| #1 \right\|} 

\newcommand{\Ip}[2]{\left\langle #1, \, #2\right\rangle} 

\DeclareMathOperator*{\argmin}{argmin}

\DeclareMathOperator{\grad}{grad}

\DeclareMathOperator*{\real}{Re}

\newcommand{\figref}[1]{Fig.\ \ref{#1}}

\newcommand{\Deleted}[1]{\ifdefined\WITHMARKEDCHANGES\deleted[id=SH]{#1}\else\fi}
\newcommand{\Added}[1]{\ifdefined\WITHMARKEDCHANGES\added[id=SH]{#1}\else#1\fi}
\newcommand{\Replaced}[2]{\ifdefined\WITHMARKEDCHANGES\replaced[id=SH]{#1}{#2}\else#1\fi}

\begin{document}

\title{Fast algorithms for nonlinear and constrained phase retrieval in near-field X-ray holography based on Tikhonov regularization}

\author{Simon Huhn,\authormark{1} Leon Merten Lohse,\authormark{1,2}  Jens Lucht\authormark{1} and Tim Salditt\authormark{1,*}}
\address{\authormark{1}Institut f\"ur R\"ontgenphysik, Universit\"at G\"ottingen, Friedrich-Hund-Platz 1, 37077 G\"ottingen, Germany\\
\authormark{2}Deutsches Elektronen-Synchrotron DESY, Notkestrasse 85, 22607 Hamburg, Germany}
\email{\authormark{*}tsaldit@gwdg.de}


\begin{abstract}
Based on phase retrieval, lensless coherent imaging and in particular holography offers quantitative phase and amplitude images.
This is of particular importance for spectral ranges where suitable lenses are challenging, such as for hard x-rays. 
Here, we propose a phase retrieval approach for inline x-ray holography based on Tikhonov regularization applied to the full nonlinear forward model of image formation. 
The approach can be seen as a nonlinear generalization of the well-established contrast transfer function (CTF) reconstruction method.
While similar methods have been proposed before, the current work achieves nonlinear, constrained phase retrieval at competitive computation times. 
We thus enable high-throughput imaging of optically strong objects beyond the scope of CTF. 
Using different examples of inline holograms obtained from illumination by a x-ray waveguide-source, 
we demonstrate superior image quality even for samples which do not obey the assumption of a weakly varying phase.
Since the presented approach does not rely on linearization, we expect it to be well suited also for other probes such as visible light or electrons, which often exhibit strong phase interaction.
\end{abstract}

\section{Introduction}

X-ray imaging and tomography offers a set of advantages for biomedical and materials research: resolution, penetration power, high volume throughput, compatibility with hydrated specimen, and quantitative contrast  mechanisms relatively unaffected by multiple scattering.  The advantage of phase over absorption contrast for brain tomography has been realized early, since for the high photon energies $E$ of hard x-rays, the index of refraction $n(E, {\mathbf r})= 1 -\delta(E,{\mathbf r}) + i\beta(E,{\mathbf r})$ leads to dominating phase interaction compared to absorption. For low-$Z$ elements, in particular, we have $\delta/\beta \gg 1$,  and a proportionality of $\delta (\mathbf r)$ and the electron density.  Hence, it has been a major research goal over the last two decades to exploit the small phase shifts in a wavefront imparted by matter for high resolution imaging. 
Different phase-contrast methods can be used to transform these phase shifts into measurable intensities patterns by ways of interference \cite{Nugent_ContemporaryPhysics_2011,Paganin_2006}, based on different geometries and mechanisms \cite{Bravin_2012,Wen_2019}. Propagation-based imaging  only requires the self-interference of the diffracted beam behind the object and the unattenuated or weakly attenuated primary beam \cite{Salditt__2020}. It hence does not require scanning of the object or any optical element, which makes the method also dose effective. Finally, the imaging modality can cover a wide range of scales, from macroscopic (clinical) settings such as mammography \cite{TavakoliTaba_2020} down to imaging sub-cellular spatial resolution below \SI{50}{\nm} \cite{Bartels_PRL_2015} using synchrotron radiation. 

A central challenge in propagation-based imaging has been the formulation of accurate and efficient phase retrieval schemes \cite{Gureyev_Appl.Opt._2004,Nugent_Adv.Phys._2010,Cloetens_APL_1999}. In the so-called  direct-contrast regime, i.e.\@ at small propagation distances $z$ or correspondingly large Fresnel numbers, linearization in $z$  can be used to derive approximate inversion equations \cite{Paganin_J.Microsc._2002}. \Replaced{On the contrary, the present work is concerned with the holographic regime of small Fresnel numbers, which is exploited for imaging with spatial resolutions in the order of \SI{100}{\nm} or below, where such inversion methods fail \cite{Burvall_OE_2011,Salditt__2020}.}{These approaches fail, however, in the holographic regime which is exploited for high resolution imaging \cite{Burvall_OE_2011,Bartels_PRL_2015}.}
In this regime, where propagation imaging is also known as near-field holography (NFH), phase-sensitivity is high enough to 'pick up' the small differences, for example encountered in unstained soft tissues. 
\Added{This enables a wide range of applications such as virtual histology of COVID-19 lung tissue \cite{Eckermann2020VirtualHistologyCovid}, structural imaging of microscopic catalyst particles \cite{Vesely2021CatParticle}, observation of expanding cavitation bubbles \cite{Vassholz2021CavitationBubbles} or resolving neuronal structures throughout entire \emph{Drosophila melanogaster} brains \cite{Kuan2020DrosophilaBrainNFH}.}
For NFH, an inversion based on the known contrast transfer function (CTF) for optically weak objects has been proposed twenty years ago \cite{Cloetens_APL_1999,Turner_OE_2004,Zabler_RSI_2005}, using holograms data recorded at different
measurement planes and Fresnel numbers $F_1, \ldots, F_J$.
This so-called CTF phase retrieval has been extremely successful, and has found widespread use also due to its numerical simplicity and efficiency.

However, CTF fails or has limited image quality in many cases, when the basic assumption of a weakly varying phase is not fulfilled \cite{Guigay_Optik_1977,Moosmann_Opt.Express_2011, Villanueva-Perez_Opt.-Lett._2017}. 
For example, interior or exterior interfaces can result in large phase gradient which often substantially deteriorate the image quality, even if the bulk of the sample is characterized by smaller phase gradients. 
To treat data with larger phase gradients, iterative phase retrieval schemes have to be considered that overcome the limitation to the linear CTF regime. 
Such nonlinear reconstructions in NFH, using gradient descent algorithms, have been reported over a decade ago \cite{Davidoiu_Opt.Express_2011,Langer_2012}. Regularized Newton methods have also been proposed \cite{Maretzke_Opt.Express_2016}. More recently, Hagemann et al. introduced a simple scheme of alternating projections (AP), which operates on the same input data and constraint sets as the conventional CTF-based phase retrieval \cite{Hagemann_APL_2018}. 
As in far-field coherent diffractive imaging (CDI), iterative projection algorithms \cite{Gerchberg_Optik_1972, Luke_SIAMreview_2002}, which cycle between measurement and object plane, do not rely on linearization and are well suited also for phase retrieval in the NFH regime. 
In addition to the measured data, a priori constraints (prior information) on the imaged object, such as finite support, separability, or range constraints on the numerical phase values can be used \cite{Giewekemeyer_PRA_2011,Pein_OE_2016}. 
However, single distance acquisitions of extended (not compactly supported) specimen, which in practice are of particular relevance for tomography, often suffer from stagnation and noise-induced artifacts as phase retrieval becomes more strongly ill-posed in this setting \cite{Maretzke_SJoAM_2017,Maretzke_Diss}. In fact, without a compact support as in \cite{Giewekemeyer_PRA_2011,Bartels_PRL_2015}, AP-type algorithms often only yield low-quality reconstructions, unless the data acquisition is extended by longitudinal (i.e.\ parallel to the optical axis) or lateral scanning.
Moreover, iterative reconstructions are typically numerically expensive, which is prohibitive for large tomographic data sets.
In order not to bottleneck tomographic imaging pipelines, computation times for image-\emph{reconstruction} should not exceed the typical image-\emph{acquisition} times ($\lesssim \SI{1}{\s}$). On the contrary, iterative projection algorithms applied to holograms with $2000^2$ pixels typically require thousands of iterations and take \emph{minutes} to complete even on data center graphical processing units (GPUs) \cite{Hagemann_APL_2018}.

In this work, we propose an algorithm for phase retrieval \Added{in the holographic regime}, denoted as NLTikh, using Tikhonov regularization to invert the full nonlinear forward model of NFH.
The approach is a nonlinear generalization of the well-established CTF reconstruction, which no longer depends on a weak object assumption and on the other hand permits to impose support- and/or range constraints on the recovered phase-image. As an intermediate step, we also introduce a constrained CTF algorithm (cCTF) that retains the linear CTF model but adds flexibility in terms of incorporating constraints.
The proposed methods overcome the aforementioned performance issues of other iterative methods by applying state-of-the-art optimization methods for minimizing the Tikhonov functional, significantly reducing the number of required iterations compared to previous approaches. Additionally, our algorithms make use of GPUs whenever available, while retaining the ease-of-use of CTF phase retrieval from the user's perspective.

While the early phase retrieval algorithms have been reported solely in the literature without usable implementations, it has become customary to publish the source code as well \cite{Lohse_Holotomotoolbox_2020,Holotomotoolbox_repository,Langer_2021}. We follow this trend and provide Matlab implementations for the proposed methods.




\section{Phase retrieval algorithms} \label{S:algo}

The task of phase retrieval in NFH is to reconstruct the phase $\phi$ (and absorption $\mu$) from near-field intensity data $I$
measured in the detector plane: 
\begin{align}
    I = \left| \Fresnel_F (\exp( \I \phi - \mu )) \right|^2 =: \NLOp_F (\I\phi - \mu) ~.
    \label{eq:modelGeneral}
\end{align}
$\Fresnel_F(\psi) = \FT^{-1} ( \exp(-\I \xi^2 /(4\pi F)) \cdot \FT( \psi) )$ denotes Fresnel-propagation of a wave-field $\psi$ from the object plane to the detector, with the Fourier transform $\FT$ and the dimensionless Fresnel number $F$, where $F\ll 1$ corresponds to the holographic regime. The object exit wave $\psi = \exp(\I \phi - \mu)$ is given by
projections of the imaged sample's optical indices $\phi = -k \int \delta(x,y,z) \D z $ and absorption $\mu = k \int \beta(x,y,z) \D z $ ($n = 1 - \delta + \I \beta$: refractive index, $z$: optical axis, $k$: wavenumber), as long as diffraction within the object can be neglected. 
The widely used \emph{weak object approximation}, i.e.\@ the linearization of \eqref{eq:modelGeneral} with respect to the phase- and absorption images $\phi$ and $\mu$, then leads to the  contrast transfer function (CTF) model \cite{Guigay_Optik_1977},
\begin{align}
    I  \approx 1 + 2 \FT^{-1}( s_F \cdot \FT(\phi)) - 2 \FT^{-1}( c_F \cdot \FT(\mu)) =: \LinOp_F(\I \phi - \mu) ~, \label{eq:modelCTF}
\end{align}
where $s_F(\xi) = \sin(\xi^2 /(4\pi F))$ and $c_F(\xi) = \cos(\xi^2 /(4\pi F))$ denote the phase- and absorption CTFs. Direct inversion of the CTF-model, as introduced by Peter Cloetens  twenty years ago \cite{Cloetens_APL_1999}, is the standard phase retrieval method in the holographic regime today. 
It works by computing a regularized inverse of \eqref{eq:modelCTF} based on a Fourier-filter constructed from the CTF. The method makes use of either a  \emph{pure phase} or a \emph{single-material approximation}, i.e.\@ $\mu=0$ or -- more generally -- proportionality of $\mu$ and $\phi$ is assumed:
\begin{align}
    \mu/\phi = \beta/ \delta = c_{\beta/\delta} = \textup{constant} ~. \label{eq:singleMat}
\end{align}

Mathematically, CTF phase retrieval amounts to applying \emph{Tikhonov regularization} to the linearized forward model under the constraint in \eqref{eq:singleMat}, i.e.\ it can be written as
\begin{subequations}
\begin{align}
	 \phi_{\textup{CTF}} &= \argmin_{\phi \in L^2(\mR^2, \mR) } \underbrace{\sum_{j = 1}^J  \Norm{ \LinOp_{F_j} ((\I-c_{\beta/\delta})\phi) - I_j  }^2  + \Norm{ \alpha^{1/2} \cdot \FT(\phi) }^2}_{=:\LinTikhFunc(\phi)} \label{eq:CTF-a} \\
	 &= \FT^{-1} \left(\frac{ 2 \sum_{j=1}^J (s_{F_j} - c_{\beta/\delta}\cdot c_{F_j}) \cdot \FT(I_j-1) }{ \alpha +  4 \sum_{j=1}^J (s_{F_j} - c_{\beta/\delta}\cdot c_{F_j})^2 }   \right) \label{eq:CTF-b}
\end{align} \label{eq:CTF}
\end{subequations}
where the $I_j$ are the holograms measured at Fresnel numbers $F_1,\ldots,F_J$, $\norm{f}:=(\int |f(x)|^2 \D x)^{1/2}$ denotes the $L^2$-norm and the weighting function $\alpha \geq 0$ controls the strength of the regularization term $\norm{ \alpha^{1/2} \cdot \FT(\phi) }^2$ in the different spatial frequencies of $\phi$.
\Added{Note that, if regularization was omitted, i.e.\ $\alpha = 0$, the denominator in \eqref{eq:CTF-b} could become arbitrarily small for spatial frequencies where the summed CTF $\sum_{j=1}^J (s_{F_j} - c_{\beta/\delta}\cdot c_{F_j})^2$ is close to zero. This would lead to a prohibitively strong amplification of data noise in the reconstructed images.}
As is customary,
we consider a two-level regularization, exhibiting a smooth transition from a weaker penalization of low spatial frequencies to a stronger penalization beyond the first maximum of the phase-CTF:
\begin{align}
\alpha(\xi) \approx \begin{cases} \alpha_{\textup{low-freq}} &\textup{for } |\xi| < \pi (2 \bar{F} )^{1/2} \\
\alpha_{\textup{high-freq}} &\textup{for }  |\xi| > \pi (2 \bar{F} )^{1/2} \\
\end{cases} \label{eq:TwoLevelRegularization}
\end{align}
where $\bar F := J^{-1} \sum_{j=1}^J F_j$ is the mean Fresnel number of the  holograms.
We use $\alpha_{\textup{low-freq}} = 10^{-3}$ and $\alpha_{\textup{high-freq}} = 10^{-1}$ as default values, which turn out to yield good results in many settings\Added{, although the 'ideal' choice of regularization parameters is application-specific.}

In the present work, we want to build upon this successful model and achieve a generalization, rather than a replacement.    
To this end, we propose two extensions: First, we introduce a CTF-based method which is more flexible with respect to
incorporation of additional \emph{a priori constraints}, including in particular \emph{support}: $\phi|_{\mR^2\setminus \Omega} = 0$  outside some known bounded domain $\Omega \subset \mR^2$,    
and \emph{negativity}: $\phi, -\mu \leq 0$, which results from the positive definiteness of electron density.
Formally, such a \textit{constrained CTF} (cCTF) reconstruction simply amounts to restricting the search set of the minimization problem in \eqref{eq:CTF}:
\begin{align}
	\phi_{\textup{cCTF}} = \argmin_{\phi \in A}  \LinTikhFunc(\phi) \MTEXT{with} A=\{\phi \in L^2(\mR^2, \mR): \phi\textup{ satisfies constraints}\}
	\label{eq:cCTF}
\end{align}

Second, and more importantly, we want to omit the weak-object assumption underlying to CTF phase retrieval, by replacing the linearized model $\LinOp_{F}$ with the general nonlinear forward map $\NLOp_{F}$. Accordingly, we aim to reconstruct $\phi$ by solving the following minimization problem:
\begin{subequations}
\begin{align}
 \phi_{\textup{NLTikh}} &\in \argmin_{\phi \in A} \NLTikhFunc(\phi) \label{eq:NLTikh-a} \\
 \MLTEXT{with} \NLTikhFunc(\phi) &:= \sum_{j = 1}^J  \Norm{ \NLOpa{F_j}   ((\I-c_{\beta/\delta})\phi) - I_j  }^2  + \Norm{ \alpha^{1/2} \cdot \FT(\phi) }^2. \label{eq:NLTikh-b}
\end{align} \label{eq:NLTikh}
\end{subequations}

While the proposed generalizations are hence conceptually straightforward, they bring about significant challenges on the practical, algorithmic side: contrary to \eqref{eq:CTF}, the minimizer in the constrained variant \eqref{eq:cCTF} can no longer be computed by a single-step FFT-based filtering operation as given by \eqref{eq:CTF-b}. Moreover, the minimization problem in \eqref{eq:NLTikh} even becomes \emph{non-convex}, so that a global solution strategy is completely beyond reach.
In both cases, we therefore have to resort to iterative algorithms to compute the sought minimizers. 

To achieve reasonably fast termination, we employ the following methods:
\begin{enumerate}
    \item \emph{State-of-the-art optimizers:} For cCTF we employ an accelerated variant \cite{Goldstein_SIAMImaging_2014_accADMM} of the alternating direction method of multipliers (ADMM). 
    For the nonlinear case we use a (projected) gradient-descent method with adaptive Barzilai-Borwein stepsizes, non-monotone linesearch and a gradient-based stopping rule as described in \cite{Goldstein_arXiv_2014_FASTA}.
    \item \emph{Warm starting:} The NLTikh-algorithm is initialized by a linear CTF reconstruction obtained with the same parameters. This helps to avoid local minima associated with phase-wrapping and typically reduces the number of iterations required for convergence.
    \item \emph{GPU computing:} Our implementation automatically performs massively parallelized computations on a graphical processing unit (GPU) if such hardware is available.
\end{enumerate}

Notably, all of these algorithmic tweaks act behind the scenes -- from the user's perspective, our cCTF and NLTikh implementations are as easy to use as standard CTF phase retrieval.
The code is fully publicly available as part of the HoloTomoToolbox \cite{Lohse_Holotomotoolbox_2020,Holotomotoolbox_repository}.

\paragraph{Relation to existing methods:} The proposed algorithms bear similarities to several previous work in the literature. In \cite{Langer_2012,Langer_2021,Wittwer2022}, gradient-descent schemes are proposed to iteratively minimize the distance to the hologram data (or rather to  square-root-data $\sqrt{I_j}$ in \cite{Wittwer2022}) based on the nonlinear model, i.e.\ the first term on right-hand side of \eqref{eq:NLTikh-b}. Moreover, the approach in \cite{Langer_2012} also includes warm-starting from a CTF-reconstruction and the algorithm in \cite{Wittwer2022} allows to impose range- or support-constraints via projected gradient descent as in the current work. On the other hand, these previous methods both lack the ability to impose regularization as well as adaptive stepsizes to accelerate convergence, leading to as many as 10000 iterations for a single reconstruction in \cite{Wittwer2022}.
Perhaps the most closely related work to the present one is \cite{Davidoiu_Opt.Express_2011}, where nonlinear Tikhonov regularization with a gradient-penalty (corresponding to the choice $\alpha(\xi) = \textup{constant} \cdot |\xi|$ in \eqref{eq:NLTikh-b}) is proposed. However, no support- or range-constraints are enabled and the algorithm involves a computationally expensive line-search strategy. Also the regularized Newton method in \cite{Maretzke_Opt.Express_2016} is similar to the present approach, but involves nested (outer) Newton- and (inner) conjugate-gradient-iterations that spoil the computational performance.
In conclusion, we find that our NLTikh algorithm stands out from previous methods in three respects: (1) it is flexible in terms of incorporating constraints, (2) builds upon experience gained with CTF phase retrieval, in particular retaining the established regularization strategy, and (3) is algorithmically optimized to keep computational costs as low as possible.

Concerning our constrained CTF method, we note that similar ADMM-schemes have been proposed for linearized phase retrieval with total variation (TV) regularization \cite{Villanueva-Perez_Opt.-Lett._2017,Kostenko2013CTFPhaseRecWithTV}. While such a regularization strategy is somewhat more advanced than our approach, we note that it has been so far limited to the weak object regime due to the algorithmic challenges it brings about.

\section{Algorithmic details}

\subsection{ADMM for constrained CTF}
We solve \eqref{eq:cCTF} by splitting it into the uncoupled subproblems of the unconstrained CTF \eqref{eq:CTF} on the one hand and the constraints on the other hand.
This splitting enables us to use an augmented Lagrangian method, namely ADMM \cite{Combettes_2011,Luke__2020}. The resulting iterations read
\begin{subequations}
\begin{align}
    \phi_{k+1} &= \argmin_{\phi \in L^2(\mR^2, \mR) } \LinTikhFunc(\phi) + \frac{\rho}{2} \Norm{\phi - \lambda_k}^2 \label{eq:cCTF_ADMM-a} \\
    &= \FT^{-1} \left(\frac{ \rho \FT(\lambda_k)  +  2 \sum_{j=1}^J (s_{F_j} - c_{\beta/\delta}\cdot c_{F_j}) \cdot \FT(I_j-1) }{ \rho + \alpha + 4 \sum_{j=1}^J (s_{F_j} - c_{\beta/\delta}\cdot c_{F_j})^2 }   \right)  \label{eq:cCTF_ADMM-b} \\
    \psi_{k+1} &= \Pi_A(\phi_{k+1} + \lambda_k) \label{eq:cCTF_ADMM-c} \\
    \lambda_{k+1} &= \lambda_k + \phi_{k+1} - \psi_{k+1}. \label{eq:cCTF_ADMM-d}
\end{align} \label{eq:cCTF_ADMM}
\end{subequations}
Here, $\lambda_k$ and $\psi_k$ are auxiliary variables that are initialized with zeros, $\rho>0$ is a stepsize parameter and  $\Pi_A$ is the projection onto the convex constraints-set $A$ from \eqref{eq:cCTF}. If, for example, a negative-phase constraint is to be imposed, i.e.\ $A=\{\phi \in L^2(\mR^2, \mR): \phi \leq 0\}$, this projection is simply a pointwise minimum: $\Pi_A(\phi) = \min (0, \phi)$. Note that the minimization in \eqref{eq:cCTF_ADMM-a} can be efficiently solved in closed form \eqref{eq:cCTF_ADMM-b}, similar to the unconstrained CTF in \eqref{eq:CTF-b}. For faster convergence, we use an accelerated variant of the basic ADMM-scheme in \eqref{eq:cCTF_ADMM}, given by Algorithm 8 in \cite{Goldstein_SIAMImaging_2014_accADMM}. We stop the iterations automatically as soon as the relative value of so-called primal and dual residual (see \cite{Goldstein_SIAMImaging_2014_accADMM}) both drop below a tolerance of $10^{-3}$.

\paragraph{Computational costs:} Noting that the
summand $2 \sum_{j=1}^J (s_{F_j} - c_{\beta/\delta}\cdot c_{F_j}) \cdot \FT(I_j-1)$ in \eqref{eq:cCTF_ADMM-b}
can be precomputed during the initialization of the algorithm, we see that the computation of the ADMM-update \eqref{eq:cCTF_ADMM} essentially requires one forward- and one inverse Fourier transform per iteration, in addition to some less expensive pointwise arithmetic operations.

\subsection{Projected gradient descent for nonlinear Tikhonov} \label{SS:NLTikhDetails}

In contrast to the linear CTF setting, no closed form, even for the case without constraints ($A = \mR^2$), exist for equation \eqref{eq:NLTikh}. However, Fr\'echet-differentiability of the nonlinear operator $\NLOpa{F}$ allows us to compute the gradient of the nonlinear Tikhonov functional $\NLTikhFunc$ in \eqref{eq:NLTikh}. As derived in appendix \ref{sec:gradient}, it can be computed as
\begin{equation}
    \grad \NLTikhFunc(\phi) 
    = 
    2\sum_{j=1}^{J} \NLOpa{F_j, \gamma}'[ \phi]^* \left( \NLOpa{F_j,\gamma}(\phi) - I_j\right) + 2 \FT^{-1}\left(\alpha  \cdot \FT(\phi) \right), \label{eq:NLTikhFuncGradient}
\end{equation}
with the abbreviations
$\gamma := \I - c_{\beta/\delta}$, $\NLOpa{F,\gamma} := \left| \Fresnel_F (\exp(\gamma \phi)) \right|^2$ and the adjoint Fréchet derivative
\begin{equation}
        \NLOpa{F,\gamma}'[\phi]^* (I) = 2 \real\left\{ \overline{\gamma \cdot \exp(\gamma\phi)} \cdot \Fresnel^{-1}_{F} \left(  \Fresnel_{F}(\exp(\gamma \phi)) \cdot I \right)\right\}. \label{eq:NLTikhAdjointDeriv}
 \end{equation}

We determine the minimizer of \eqref{eq:NLTikh} using a projected gradient descent method. The iteration is a gradient descent step followed by the projection onto the constraints-set $A$:
\begin{align}
   \phi_{k+1} &= \Pi_A \left( \phi_k - \tau_k \grad \NLTikhFunc(\phi_k) \right).
    \label{eq:pdg}
\end{align}
Without constraints, this reduces to the well-known gradient descent method.
Convergence of the algorithm crucially depends on the chosen stepsizes $\tau_k > 0$: small $\tau_k$ typically lead to slow convergence, whereas overly large stepsizes may cause the method to fail to decrease the $\NLTikhFunc(\phi)$ altogether.
We use the popular stepsize-rule proposed by Barzilai and Borwein \cite{BarzilaiBorwein1988Stepsizes}, 
\begin{align}
    \tau_k = \begin{cases}
    \frac{\ip{\phi_k - \phi_{k-1}}{\grad \NLTikhFunc(\phi_k) - \grad \NLTikhFunc(\phi_{k-1})}}{ \norm{\grad \NLTikhFunc(\phi_k) - \grad \NLTikhFunc(\phi_{k-1})}^2 } &\textup{for odd }k \\
    \frac{\norm{\phi_k - \phi_{k-1}}^2}{ \ip{\phi_k - \phi_{k-1}}{\grad \NLTikhFunc(\phi_k) - \grad \NLTikhFunc(\phi_{k-1})} } &\textup{for even }k
    \end{cases}
    \label{eq:BarzilaiBorweinStepsizes}
\end{align}
($\ip{\cdot}{\cdot}$: $L^2$-inner product) combined with a non-monotone line-search \cite{BirginEtAl2000NonmonotoneLinesearch}, as detailed in \cite{Goldstein_arXiv_2014_FASTA}.

We note that, by construction, the NLTikh-method reduces to CTF-phase retrieval in the linear weak object limit $\phi \to 0$. On the other hand, a CTF-reconstruction is numerically both more stable and less expensive to compute (also in a constrained setting!). We exploit this fact by using a CTF-phase as an initial guess for the NLTikh-algorithm, i.e.\ $\phi_0 = \phi_{\textup{cCTF}}$.

The iterations are automatically stopped as soon as the (relative) residual gradient,
\begin{align} 
R_{\textup{grad}}(\phi_k) = \frac{\norm{ \grad  \NLTikhFunc(\phi_k)}}{\norm{ \grad  \NLTikhFunc(0)}},
\label{eq:ResidualGradient}
\end{align}
drops below a threshold of $10^{-3}$. The reasoning is as follows: as the Tikhonov functional $\NLTikhFunc$ is continuously differentiable, the sought minimum is characterized by $\grad  \NLTikhFunc(\phi_{\textup{NLTikh}} ) = 0$ and $\norm{ \grad  \NLTikhFunc(\phi)}$ must tend to zero when the iterates $\phi_k$ converge to $\phi_{\textup{NLTikh}}$. In this sense, $R_{\textup{grad}}$ measures proximity to the Tikhonov minimizer, i.e.\ provides a convergence metric.

\paragraph{Stabilization of high frequencies:} The linear CTF model is diagonal in Fourier space, i.e.\ the reconstruction of a certain spatial frequency only depends on the Fourier-mode of the same frequency in the hologram data $I_1, \ldots, I_J$. In particular, this means that low-frequency data-errors may not cause high-frequency artifacts in the recontruction and vice-verser. Switching from the CTF- to the nonlinear model $\LinOp_{F_j} \to \NLOpa{F_j}$ does \emph{not} preserve this property. Hence, low-frequency background variations in the holograms, as often caused by imperfect flat-field correction, may result in oscillatory artifacts when reconstructing nonlinearly. We find that this undesirable effect can be effectively suppressed by adding a third level to the regularization defined in \eqref{eq:TwoLevelRegularization}: 

\begin{align}
\alpha(\xi) \approx \begin{cases} \alpha_{\textup{low-freq}} &\textup{for } |\xi| < \pi (2 \bar{F} )^{1/2} \\
\alpha_{\textup{high-freq}} &\textup{for }  \pi (2 \bar{F} )^{1/2} < |\xi| < \pi D \bar{F} \\
\alpha_{\textup{beyond-NA}} &\textup{for } |\xi| > \pi D \bar{F} \\
\end{cases} \label{eq:ThreeLevelRegularization}
\end{align}
where $D$ denotes the aspect length of the detector and  $\alpha_{\textup{beyond-NA}}$ is large (default: $\alpha_{\textup{beyond-NA}}= 2J$). 
As the naming suggests it, the modification enforces a higher penalization for spatial frequencies that scatter under angles exceeding the numerical aperture (NA) of the imaging system, as defined by the field-of-view covered by the detector. Note that the corresponding high-frequency Fourier-modes are thus not sufficiently represented in the acquired hologram data anyway so that these could not be stably recovered \cite{Maretzke_IP_2018}. Thus, nothing is really lost by damping out these frequencies, while stability is gained on the algorithmic side.

\paragraph{Computational costs:} Up to computationally inexpensive pointwise arithmetic operations, evaluating one projected gradient-descent step \eqref{eq:pdg} essentially requires $2J$ forward- and $J$ backward Fresnel propagations according to equations \eqref{eq:NLTikhFuncGradient} and \eqref{eq:NLTikhAdjointDeriv}. Consequently, the computational costs of an NLTikh-iteration is comparable to that of typical alternating-projection iterations. Note that neither the stepsize-rule \eqref{eq:BarzilaiBorweinStepsizes} nor checking the stopping-criterion \eqref{eq:ResidualGradient} requires significant computational effort, as $\grad  \NLTikhFunc(\phi_k)$ must be computed in every iteration anyway.


\section{Performance tests and comparison to existing methods}
The algorithms were tested with NFH data measured at the G\"ottingen instrument for nano imaging with x-rays~(GINIX), installed at the P10 beamline of the PETRA~III storage ring (DESY, Hamburg), at a photon energy of \SI{8}{\keV} (Fig.\ref{fig:spheres}, Fig.\ref{fig:sedimentation}) and \SI{13.8}{\keV} (Fig.\ref{fig:golgi}). An x-ray waveguide was used as a spatial and coherence filter of the undulator radiation which was focused onto the waveguide by a Kirkpatrick-Baez (KB) mirror \cite{Salditt_JSR_2015}. The exit of the waveguide forms a quasi-point source for holographic illumination.   
The geometrically magnified holograms of size $2048\times 2048$ pixels were recorded by two different fiber coupled scintillator sCMOS detectors (Photonic Science and Hamamatsu) 
each with pixel sizes of \SI{6.5}{\mu m}, placed approximately at $z_{02}\approx \SI{5}{\m}$ behind the KB focus. \Added{All holograms considered as test data in this work are given by dark- and flat-field-corrected detector images.}
\Added{The experimental setup parameters for the test cases in the subsequent sections are detailed in table \ref{tab:Parameters}.}

\begin{table}
\begin{center}
\begin{tabular}{  l | l  l  l  }
  & Spheres ($\S$\ref{SS:PolySpheres}) &  Hippocampus ($\S$\ref{SS:GolgiCox}) & Capillary ($\S$\ref{SS:Capillary}) \\
 \hline
 X-ray energy & \SI{8.0}{\keV} & \SI{13.8}{\keV} & \SI{8.0}{\keV} \\
 Image size (pixels) & $2048 \times 2048$ & $2048 \times 2048$ & $2048 \times 2048$   \\
 Physical pixel size & \SI{6.5}{\mu m} & \SI{6.5}{\mu m} & \SI{6.5}{\mu m}   \\
 Source to detector & \SI{5.178}{m} & \SI{5.040}{m} & \SI{5.047}{m}   \\
  Source to sample (mm) & 156, 158, 166, 187 & 134, 138, 147, 156 & 60.7  \\  
 Geometric magnification & 33.1 & 37.7 & 83.1 \\  
 Effective pixel size & \SI{196}{nm} & \SI{172}{nm} & \SI{78.2}{nm}   \\
 Fresnel numbers $(10^{-3})$ & $ 1.59, 1.57, 1.49, 1.33$ & $2.47,  2.39, 2.25, 2.12$ & $0.650$ \\  
\end{tabular}
\end{center}
\caption{\Added{Imaging setup parameters for the phase reconstructions considered in sections \ref{SS:PolySpheres} to \ref{SS:Capillary}. In the case of measurements at multiple source-sample-distances, the holograms have been rescaled to a single geometric magnification. The Fresnel numbers are stated using the effective pixel size as a lateral reference length-scale.}} \label{tab:Parameters}
\end{table}

\subsection{Test structure of polysysterene microspheres} \label{SS:PolySpheres}
\begin{figure}[htbp]
\centering
\includegraphics[width=\linewidth]{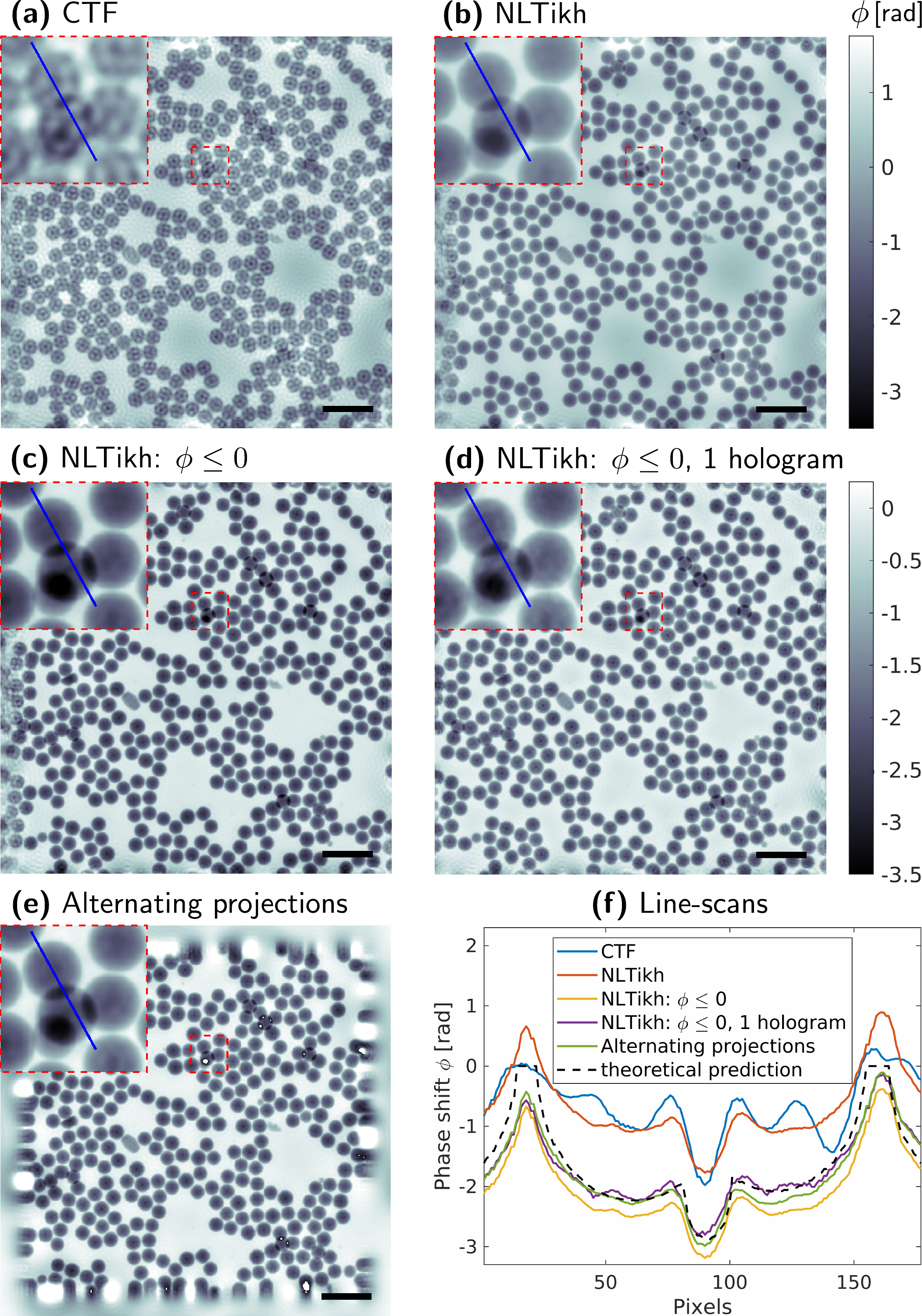}
\caption{Reconstruction of a test object composed of \SI{15}{\mu m}-sized polysterene spheres. The images plot the phases obtained by different algorithms:
\textbf{(a)} CTF phase retrieval. \textbf{(b)} The proposed NLTikh-algorithm. \textbf{(c)} Like (b) but with an additional negative-phase constraint. \textbf{(d)} Like (c) but using only one of the four holograms.
\textbf{(e)} Alternating projections (AP) result taken from \cite{Hagemann_APL_2018}.
\textbf{(f)} Line-scans through the reconstructions along the blue lines. \Added{The black-dashed line shows the theoretical phase shifts if ideal spheres and a literature value $\delta = 3.673\cdot10^{-6}$ for polystyrene are assumed.}
\Deleted{Computation times were (a) \SI{0.41}{\s}, (b) \SI{1.8}{\s} (38 gradient-descent iterations), (c) \SI{2.3}{\s} (47 iterations) and (d) \SI{2.0}{\s} (86 iterations), respectively.}
Insets in the top-left corner show zooms of the red-dashed regions. Scale bars denote \SI{50}{\mu m}.}
\label{fig:spheres}
\end{figure}

First, we demonstrate the performance of nonlinear phase retrieval using a test object of polystyrene spheres with diameter \SI{15}{\mu m} \cite{Hagemann_APL_2018}. Holograms were acquired at four different defocus distances \Replaced{(see table \ref{tab:Parameters} for detailed parameters)}{(corresponding to Fresnel numbers $F_1 = 1.333 \cdot 10^{-3}, F_2 = 1.494 \cdot 10^{-3}, F_3 = 1.570 \cdot 10^{-3}, F_4 = 1.590 \cdot 10^{-3}$ for a pixel-size length-scale)}, as in standard linearized (CTF-based) phase retrieval to compensate for minima in the CTF \cite{Cloetens_APL_1999,Zabler_RSI_2005}.
The data set has been previously considered to demonstrate the advantage of iterative phase retrieval with an alternating projections (AP) algorithm over CTF \cite{Hagemann_APL_2018}.
Figure \ref{fig:spheres} shows a comparison of reconstructions obtained by (a) standard (unconstrained) CTF, the proposed NLTikh algorithm (subfigures (b), (c), (d)) as well as (e) the AP-result from \cite{Hagemann_APL_2018}. The same parameters, $c_{\beta/\delta} = 0$ (pure phase object) and $\alpha_{\textup{low-freq}} = 10^{-3}$, $\alpha_{\textup{high-freq}} = 10^{-1}$, were used for all CTF- and NLTikh-reconstructions. \Added{Table \ref{tab:ComputationTimes} gives an overview over the observed computation times for the different reconstrution methods.}

As noted already in \cite{Hagemann_APL_2018}, the CTF result shows significant artifacts from the large phase gradients introduced by the spheres, whereas NLTikh and AP yield visually artifact-free images of similar quality. \Replaced{This is confirmed by the line-scans in \figref{fig:spheres}(f) where NLTikh and AP both show good agreement with the theoretically predicted phase profiles for $15 \, \mu \textup m$-sized polystyrene spheres (assuming the literature value $\delta = 3.673\cdot 10^{-6}$ \cite{Henke_DeltaBetaTables_1993} for the refractive index of $\textup C_8 \textup H _8$ with density $1.05\,\textup g / \textup{cm}^{3}$).}{(see also the line-scans in Fig.\ \ref{fig:spheres}(f)).} However, while the AP-reconstruction took 2500 iterations to complete \cite{Hagemann_APL_2018}, corresponding to minutes of computations on typical GPUs, the NLTikh algorithm requires only about 50 iterations to converge, thus completing within about two seconds on our benchmark workstation\Deleted{ (all GPU-computations of this work were performed in double-precision floating-point arithmetics on an NVIDIA Quadro RTX 6000 using Matlab R2020a)}. 
Moreover, Fig.\ \ref{fig:spheres}(d) demonstrates that even an NLTikh-reconstruction from a single hologram results in an acceptable image quality, apart from an expectable increase of noise compared to the four distance setting.
Finally, a comparison of Figs.\ \ref{fig:spheres} (b) and (c) shows that the ability of NLTikh to impose negativity of the reconstructed phases helps to effectively suppress low-frequency variations of the background and thus achieve quantitatively correct phases -- without a relevant increase in computational costs.

\begin{table}
\begin{center}
\begin{tabular}{  l | l l l l l }
  & CTF & NLTikh & NLTikh$_{\phi\leq0}$ & NLTikh$_{\phi\leq0}^{\textup{1 hologram}}$ & AP from \cite{Hagemann_APL_2018} \\
 \hline
 Number of iterations & / & 38 & 47 & 86 & 2500 \\
 Computation time (s) & 0.41 & 1.8 & 2.3 & 2.0 & 83.1  \\
\end{tabular}
\end{center}
\caption{\Added{Computation times for phase reconstruction of the polystyrene-microspheres data set using different algorithms. See \figref{fig:spheres}(a)-(e) for the corresponding reconstructed images, respectively. All GPU-computations were performed in double-precision floating-point arithmetics on an NVIDIA Quadro RTX 6000 using Matlab R2020a.}} \label{tab:ComputationTimes}
\end{table}

\subsection{Convergence experiments} \label{SS:Convergence}

For an exemplary assessment of convergence of the NLTikh algorithm, we consider the reconstruction setting in Fig.\ \ref{fig:spheres} (c), that includes both constraints and nonlinearity. First, we compute a (supposedly) fully converged reference solution $\phi_{\textup{ref}}$ by performing 1000 iterations of the proposed projected gradient descent scheme. We then compute the gradient residual $R_{\textup{grad}}$ defined in \eqref{eq:ResidualGradient} as well as the residual of the Tikhonov functional relative to $\phi_{\textup{ref}}$:
\begin{align} 
R_{\textup{value}}(\phi_k) = \frac{\NLTikhFunc(\phi_k)- \NLTikhFunc(\phi_{\textup{ref}})}{ \NLTikhFunc(0)-\NLTikhFunc(\phi_{\textup{ref}})}
\label{eq:ResidualFunctional}
\end{align}
Note that the quantity $R_{\textup{value}}$ can only be computed if $\phi_{\textup{ref}}$ is known, which is not the case in practical reconstruction settings. For that reason, the automatic stopping rule for the gradient-descent iterations in section~\ref{SS:NLTikhDetails} is based on $R_{\textup{grad}}$ rather than $R_{\textup{value}}$.
\begin{figure}[htbp]
\centering
\includegraphics[width=\linewidth]{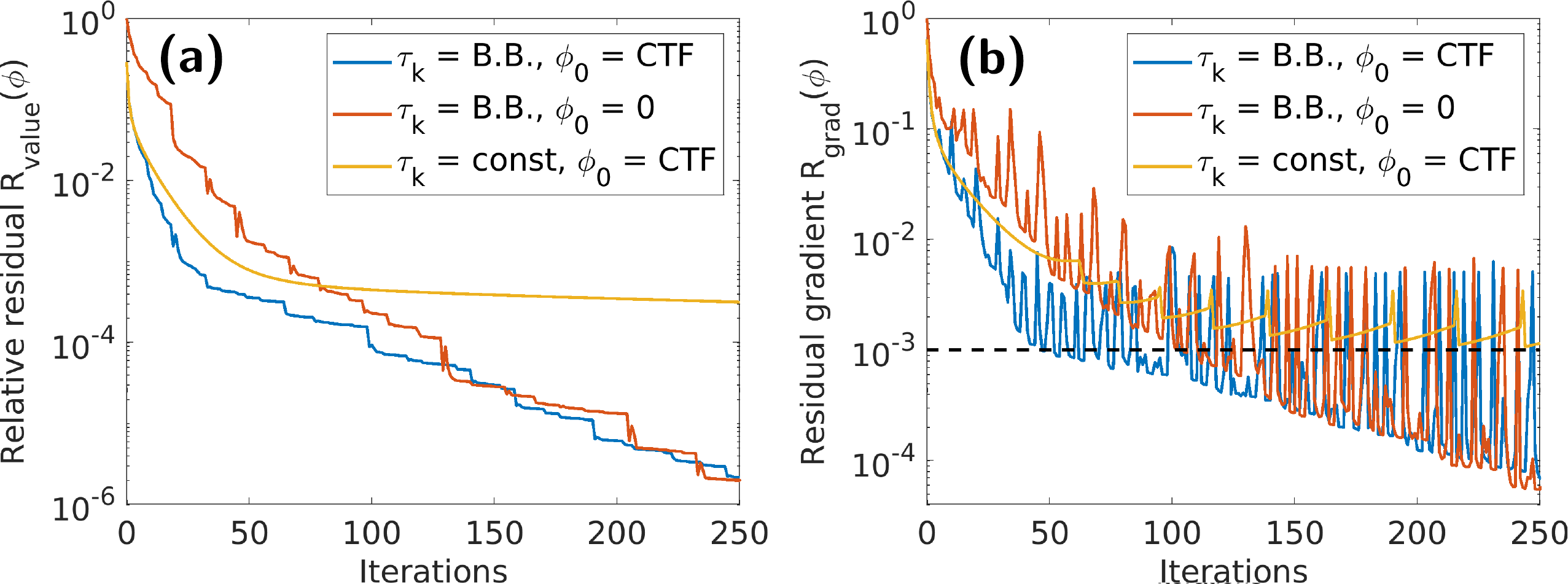}
\caption{Convergence tests for NLTikh: The plots show the residuals defined in equations \textbf{(a)} \eqref{eq:ResidualFunctional} and \textbf{(b)} \eqref{eq:ResidualGradient}, respectively, for the reconstruction setting in Fig.\ \ref{fig:spheres} (c). Blue lines: the proposed algorithm. Red: same as (a) but without warm start, i.e.\ the algorithm is initialized with $\phi_0=0$ instead of a CTF-reconstruction. Yellow: same as (a) but with constant- instead of Barzilai-Borwein stepsizes $\tau_k$. By default, the iterations are stopped when $R_{\textup{grad}}$ drops below $10^{-3}$, as marked by the black-dashed line.}
\label{fig:convergence}
\end{figure}

The convergence test is carried out both for the proposed Barzilai-Borwein stepsize-rule as well as for a constant stepsize for comparison. Moreover, to motivate our warm-starting approach, we compare reconstructions initialized with a CTF-reconstruction (as proposed) to the convergence behavior if the algorithm is initialized with zeros, $\phi_0 = 0$. Results for $R_{\textup{value}}$ and $R_{\textup{grad}}$ are plotted in Figs.\ \ref{fig:convergence} (a) and (b), respectively. The following observations can be made:
\begin{itemize}
    \item Barzilai-Borwein stepsizes yield considerably faster convergence than constant stepsizes.
    \item While the gradient-residual $R_{\textup{grad}}(\phi_k)$ may vary strongly between subsequent iterations, its overall decrease is comparable to that of $R_{\textup{value}}(\phi_k)$ for all three test cases. This emphasizes the validity of the proposed stopping criterion based on the metric $R_{\textup{grad}}$.
    \item Although the difference levels out asymptotically, the warm-start strategy accelerates initial convergence. In particular, it takes only $47$ iterations to reach the stopping criterion $R_{\textup{grad}}(\phi_k)< 10^{-3}$ with warm start compared to $105$ iterations if  initialized with $\phi_0=0$.
\end{itemize}
Overall, the convergence experiments thus show that our algorithmic optimizations to accelerate the projected gradient-descent scheme indeed pay off in the considered example.

\subsection{Golgi-Cox stained hippocampus} \label{SS:GolgiCox}

Next, we tested the nonlinear Tikhonov algorithm on a biological sample, notably a Golgi-Cox stained brain slice of 
mouse hippocampus embedded in EPON resin, which was already investigated before by CTF \cite{Toepperwien_Proc.SPIE_2016a,Toepperwien_Diss}
and the AP-algorithm \cite{Hagemann_APL_2018}.
 \Added{The imaging setup is detailed in table \ref{tab:Parameters}.}
The Golgi-Cox stain is a classical stain for brain sections, which visualizes
individual neurons by seemingly stochastic all-or-nothing events based on precipitation of silver along the neuron's neurites. 
The metalized neurites then stand out in strong contrast with respect to the neuropil infiltrated by EPON resin, resulting in strongly varying phases. 
Hence, the assumption neither of a weak object, nor of a homogeneous (single material) object are strictly met.

\begin{figure}[htbp]
\centering
\includegraphics[width=\linewidth]{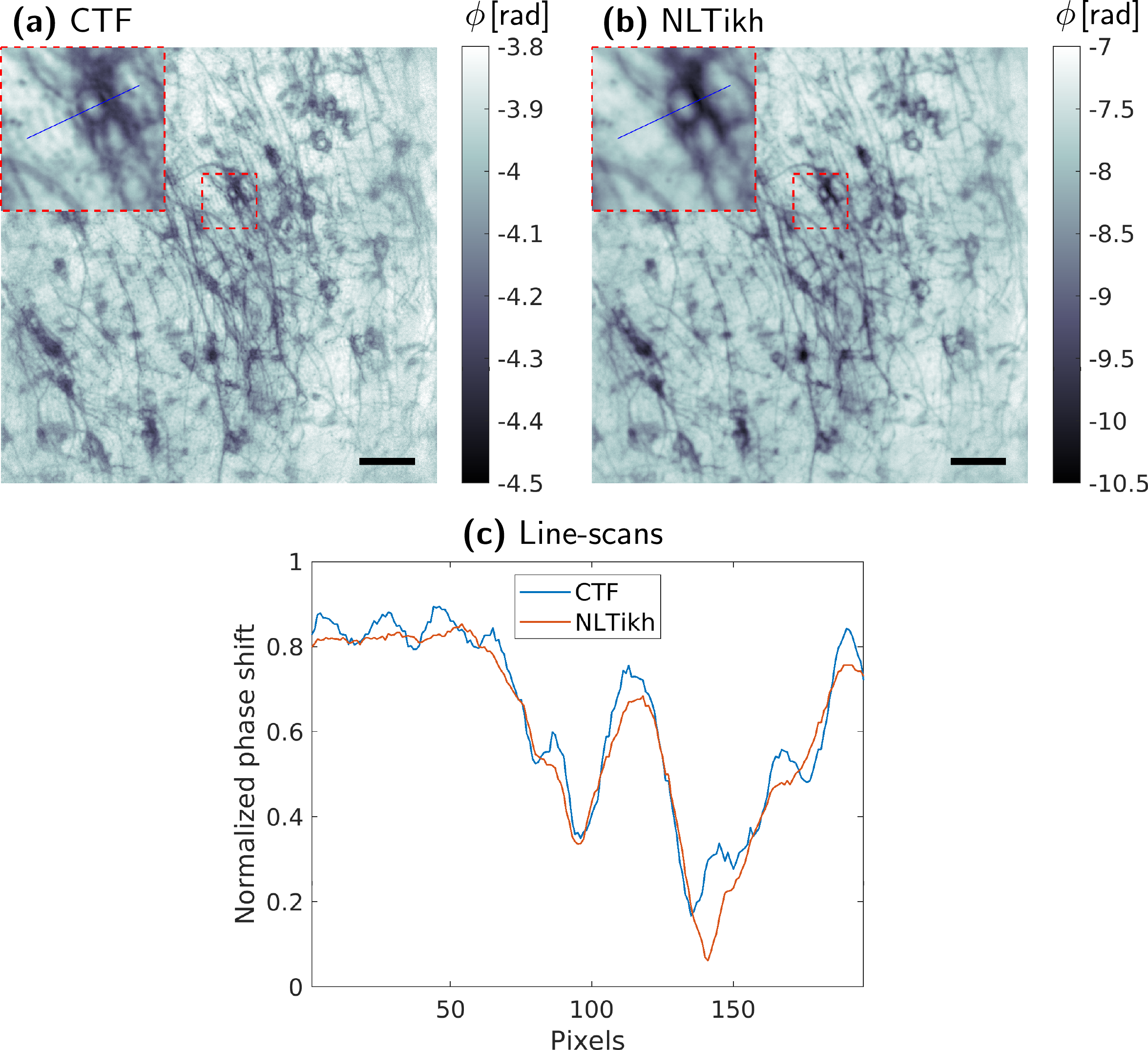}
\caption{Reconstruction mouse hippocampus tissue with Golgi-Cox silver-stained neurons. The images plot the phases obtained by different algorithms:
\textbf{(a)} (unconstrained) CTF-based phase retrieval. \textbf{(b)} The proposed NLTikh algorithm.
\textbf{(c)} Line-scans through the reconstructions along the blue lines. To improve comparability, the extracted phases are expressed relatively to the colorbar ranges in (a) and (b), respectively. 
Computation times were (a) \SI{0.45}{\s} and (b) \SI{0.90}{\s} (18 iterations).
Insets in the top-left corner show zoom-images of the red-dashed regions. Scale bars are \SI{50}{\mu m}.}
\label{fig:golgi}
\end{figure}

\figref{fig:golgi} shows reconstructions of the four-hologram data set 
\Deleted{(Fresnel numbers: $F_1 = 2.120 \cdot 10^{-3}, F_2 = 2.248 \cdot 10^{-3}, F_3 = 2.394 \cdot 10^{-3}, F_4 = 2.474 \cdot 10^{-3}$)}
obtained by (a) CTF phase retrieval, and (b) the nonlinear Tikhonov algorithm. In both cases, the same regularization parameters ($\alpha_{\textup{low-freq}} = 10^{-3}$, $\alpha_{\textup{high-freq}} = 10^{-1}$) and $c_{\beta/\delta}=0.1$ were used without additional constraints. 
Visual comparison of the reconstructed projections shows significantly less artifacts for NLTikh compared to CTF. 
In particular, the flawful wavy background at high spatial frequency, visible as oscillations of the CTF line-scan in \figref{fig:golgi}(c), is absent.
Furthermore, the NLTikh-reconstruction exhibits a much larger range of reconstructed phases compared to CTF, which is physically plausible for the high-contrast metal-stained neurons. 

\subsection{Sedimented silicon spheres in a capillary} \label{SS:Capillary}

For a final test case, we investigate a setting which also often encountered in practice, where a strong phase gradient at the edges of the object has to be imaged along with significantly weaker phase-variations in the interior. 
Examples are many: glass capillaries with tissue punches, single particles in air, trimmed specimens of material science. 
Here we consider silicon spheres of diameter \SI{2}{\mu m} in a capillary. The tomographic data set \Replaced{consists of 726 holograms acquired at a single source-to-sample distance (Fresnel number $F = 6.499\cdot 10^{-4}$) and equidistant incident angles covering a range of 180 degrees, see table~\ref{tab:Parameters} for details. It}{, consisting of a single hologram per incidence angle at Fresnel number $F = 6.499\cdot 10^{-4}$,} has been recorded as part of a sedimentation experiment where the 3D-motion of the single spheres is  tracked over time \cite{Ruhlandt_Diss}. The considered tomogram shows the final, fully sedimented state of the sample.

\begin{figure}[htbp]
\centering
\includegraphics[width=\linewidth]{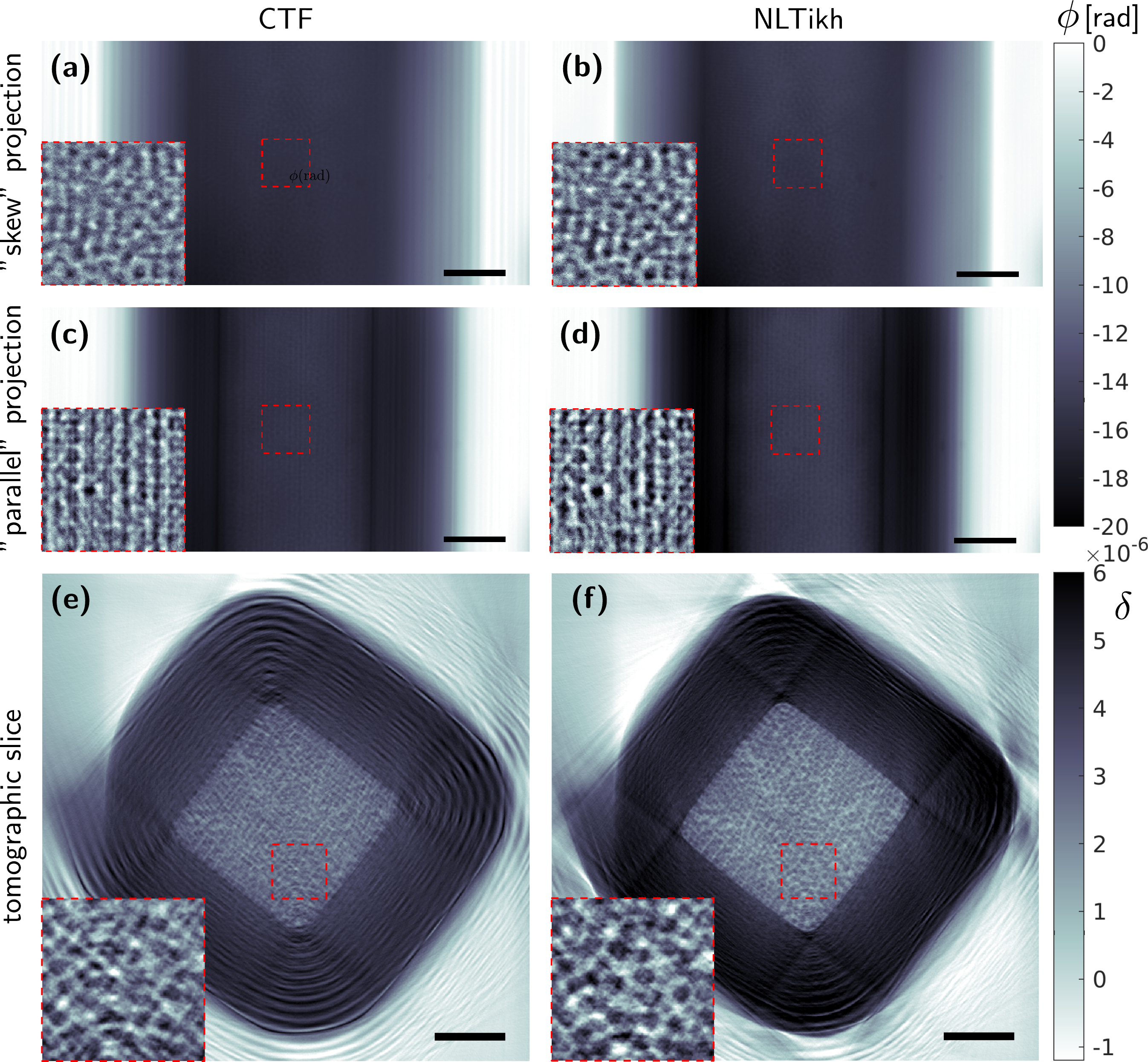}
\caption{Reconstruction of sedimented silicon spheres in a glass capillary. The left and right column show phases obtained by CTF and NLTikh, respectively.  
\textbf{(a)}, \textbf{(b)}: Projections reconstructed for skew incidence direction of the x-rays with respect to the capillary boundaries. \textbf{(c)}, \textbf{(d)}: Projections for an almost parallel incidence. \textbf{(e)}, \textbf{(f)} \Replaced{Central tomographic slices computed from the reconstructed phase projections for all 726 incident angles using filtered back-projection.}{Central slice through tomographic reconstructions.}
Insets in the bottom-left corner show zoom-images of the red-dashed regions, respectively. Scale bars denote \SI{10}{\mu m}.}
\label{fig:sedimentation}
\end{figure}

We reconstruct the \Replaced{holograms for each incidence angle using}{data with} CTF and NLTikh, imposing the $\beta/\delta$-ratio of the glass capillary ($\textup{SiO}_2$) at a photon energy of \SI{8}{\keV}, $c_{\beta/\delta} = 0.0135$ according to \cite{Henke_DeltaBetaTables_1993}, as well as negativity of the recovered phases. Regularization parameters are $\alpha_{\textup{low-freq}} = 10^{-5}$ and $\alpha_{\textup{low-freq}} = 10^{-1}$. Results are shown in Fig.\ \ref{fig:sedimentation}.

The capillary has an approximately square cross-section. Accordingly, the severeness of the phase gradients at the capillary-boundary depends on the incidence direction of the x-rays: when the latter is skew with respect to the sides of the capillary, only moderate phase-variations occur, whereas strongly varying phases result in the case of (almost) parallel incidence. Fig.\ \ref{fig:sedimentation} (a), (b) and (c), (d) shows exemplary projections reconstructed for skew and parallel settings, respectively. We observe that the CTF-reconstructions suffer from fringe-artifacts in both cases, which also distort the sphere-structures in the interior of the capillary. For NLTikh, such artifacts still occur in the parallel-incidence setting, whereas the skew reconstruction is practically artifact-free -- despite the strong total phase-shifts up to $\approx \SI{20}{rad}$. As revealed by the tomographic slices (computed \Replaced{by applying filtered back-projection to the recontructed phase images for all tomographic angles)}{using filtered back-projection}) in Fig.\ \ref{fig:sedimentation} (e), (f), our nonlinear NLTikh method overall achieves an improved, yet still imperfect reconstruction quality compared to CTF, yielding a clearer but still not artifact-free image of the silicon spheres' packing in the capillary.

\section{Summary and Outlook}

In summary, we have presented a nonlinear Tikhonov phase retrieval algorithm \Added{for near-field holography (NFH)}, to cope with optically non-weak objects beyond the scope of the established linear CTF model. Since the NLTikh approach comprises standard CTF phase retrieval as a limiting case, it can be regarded as a nonlinear generalization of the CTF approach.
We have validated NLTikh for experimental data, including test objects and 'real samples', recorded under relevant conditions of nanoscale holographic x-ray imaging\Added{, both in 2D and in a tomographic 3D setting}.
As expected by the algorithmic design, which is based on the full nonlinear model of NFH, the method is found capable to reconstruct objects with moderately strongly varying phase shifts and outperforms CTF-based methods in terms of image-quality in such settings.
As a further option, NLTikh enables imposing additional support and range constraints -- like the proposed constrained CTF (cCTF) algorithm for the weak object regime. 
This is of particular interest if hologram data is available only for a single distance, where imposing additional a priori knowledge may compensate for the lack of data richness \cite{Maretzke_SJoAM_2017,Maretzke_Diss}.
We emphasize that both the ability to reconstruct optically non-weak objects and to exploit additional a priori constraints are also provided by previously considered methods, such as alternating projection schemes (AP). 
The distinctive feature of our NLTikh algorithm, however, lies in the demonstrated superior numerical efficiency, rendering nonlinear, constrained image-reconstruction applicable for large-scale tomographic data sets with thousands of high-resolution holograms to be processed.

Despite the merits of the presented approach, one should also note its limitations. While not being strictly limited to the weak object regime like CTF, NLTikh still relies on linearity to a certain degree: it uses a CTF-reconstruction as an initial guess and the iterations are based on derivatives of the nonlinear model, i.e.\ on local linear approximations. Due to non-convexity, convergence is thus only guaranteed if the true phase is sufficiently close to the linear CTF-result, i.e.\ nonlinearity must not be too strong.
For very strong objects, as considered in section \ref{SS:Capillary}, the initial guess may be too far off for the gradient-descent scheme to find the global minimum of the Tikhonov functional.
In such a setting, AP and related methods like RAAR \cite{Luke_IP_2004} may perform significantly better than NLTikh, as they are originally designed for highly non-convex image-reconstruction problems like CDI.
A second limitation of our algorithm lies in the assumption of a homogeneous (single material) object. Like the generalizations of CTF phase retrieval made in this work, lifting this assumption is conceptually simple: just optimize the Tikhonov functional in \eqref{eq:CTF}, \eqref{eq:cCTF} or \eqref{eq:NLTikh} for both parameters $\phi$ and $\mu$, omitting the coupling constant $c_{\beta/\delta}$. Yet, achieving accurate image reconstructions in a numerically stable and efficient manner by such an approach is expected to constitute a severe algorithmic challenge, that might be addressed in future research.
\Added{Owing to the advent of energy-resolving X-ray detectors, another interesting research direction might be to extend NLTikh to settings with holograms acquired at different X-ray energies, similar to what has been demonstrated in the direct-contrast regime \cite{Schaff2020SpectralDirectContrastPCI,Ghani2021DualEnergyDirectContrastPCI}.}



\appendix

\section{Derivation of the gradient}
\label{sec:gradient}

The nonlinear forward model $\NLOpa{F,\gamma}$ in \eqref{eq:NLTikhFuncGradient} is Fréchet-differentiable with derivative (see \cite{Maretzke_Opt.Express_2016})
\begin{equation}
        \NLOpa{F,\gamma}'[\phi] (\psi) = 2 \real\left\{ \gamma \cdot \overline{\Fresnel_{F}(\exp(\gamma \phi))} \cdot \Fresnel_{F}(\exp(\gamma \phi) \cdot \psi )\right\}.
\end{equation}
In order to compute the adjoint derivative, we consider the $L^2$-inner product of $\NLOpa{F,\gamma}'[\phi](\psi)$ with an intensity $I$ for arbitrary real-valued functions $\phi$, $\psi$, and $I$:
\begin{equation}
\begin{split}
    \left\langle \NLOpa{F,\gamma}'[\phi](\psi), I \right\rangle 
    &= \left\langle 2 \real \left\{ \gamma \cdot \overline{\Fresnel_F(\exp(\gamma \phi))} \cdot \Fresnel_F(\exp(\gamma \phi) \cdot \psi) \right\}, I \right\rangle \\
    &= 2 \real \left\langle \gamma \cdot \overline{\Fresnel_F(\exp(\gamma \phi))} \cdot \Fresnel_F(\exp(\gamma \phi) \cdot \psi) , I \right\rangle \\
    &= 2 \real \left\langle \Fresnel_F(\exp(\gamma \phi) \cdot \psi) , \overline \gamma \cdot {\Fresnel_F(\exp(\gamma \phi))} \cdot I \right\rangle \\
    &= 2 \real \left\langle \Fresnel_F^{-1} \Fresnel_F(\exp(\gamma \phi) \cdot \psi) , \overline \gamma \cdot \Fresnel_F^{-1}\left({\Fresnel_F(\exp(\gamma \phi))} \cdot I\right) \right\rangle \\
    &= 2 \real \left\langle  \psi , \overline{ \gamma\cdot \exp(\gamma \phi)} \cdot \Fresnel_F^{-1}\left({\Fresnel_F(\exp(\gamma \phi))} \cdot I\right) \right\rangle \\
    &= \left\langle  \psi , 2 \real \left\{ \overline{ \gamma\cdot \exp(\gamma \phi)} \cdot \Fresnel_F^{-1}\left({\Fresnel_F(\exp(\gamma \phi))} \cdot I\right)\right\} \right\rangle,
    \end{split} \label{eq:adjoint-1}
\end{equation}
where we used that $\Fresnel_F$ is unitary as well as real-valuedness of $\phi$, $\psi$, and $I$.
By the defining property of the adjoint, we also have that
\begin{equation}
    \left\langle \NLOpa{F,\gamma}'[\phi](\psi), I \right\rangle = \left\langle \psi, \NLOpa{F,\gamma}'[\phi]^\ast (I) \right\rangle. \label{eq:adjoint-2}
\end{equation}
Since \eqref{eq:adjoint-1} and \eqref{eq:adjoint-2} hold for arbitrary functions $\phi,\psi,I$, we may thus read off that
\begin{equation}
        \NLOpa{F,\gamma}'[\phi]^\ast( I ) = 2  \real\left\{ \overline{\gamma \cdot \exp(\gamma\phi)} \cdot \Fresnel^{-1}_{F} \left(  \Fresnel_{F}(\exp(\gamma \phi)) \cdot I \right)\right\}.
\end{equation}

Now we proceed to computing the gradient of the nonlinear Tikhonov functional $\NLTikhFunc$ from \eqref{eq:NLTikh}. Noting that the Fr\'echet-derivative of $G(f):= \norm{f}^2 = \ip{f}{f}$ is given by $G'[f](h) = 2\ip{f}{h}$ and that $\NLTikhFunc(\phi) = \sum_{j = 1}^J  G( \NLOpa{F_j}   ((\I-c_{\beta/\delta})\phi) - I_j ) + G(\alpha^{1/2}\cdot\FT(\phi))$, we obtain by application of the sum- and chain-rule for Fr\'echet-derivatives
\begin{align}
    \ip{\grad\NLTikhFunc(\phi)}{h} &= \NLTikhFunc'[\phi](h) \nonumber \\
    &=\sum_{j = 1}^J G'\left[\NLOpa{F_j}(\phi)-I_j\right]  \left(\NLOpa{F_j}'[\phi](h)\right) + G'\left[\alpha^{1/2}\cdot\FT(\phi)\right] \left(\alpha^{1/2}\cdot\FT(h)\right) \nonumber \\
    &=2 \sum_{j = 1}^J \Ip{\NLOpa{F_j}(\phi)-I_j}{\NLOpa{F_j}'[\phi](h)} +   2 \Ip{\alpha^{1/2}\cdot\FT(\phi)}{\alpha^{1/2}\cdot\FT(h)} \nonumber \\
    &= \Ip{2 \sum_{j = 1}^J \NLOpa{F_j}'[\phi]^\ast \left( \NLOpa{F_j}(\phi)-I_j \right) + 2 \FT^{-1} \left( \alpha\cdot\FT(\phi) \right)}{h} \, . \label{eq:app-gradient-1}
\end{align}
In the final step, we used that $\phi \mapsto \alpha^{1/2} \cdot \FT(\phi)$ is linear with adjoint $\psi \mapsto \FT^{-1} ( \alpha^{1/2} \cdot \psi)$ together with linearity of the inner product.
Since equation \eqref{eq:app-gradient-1} holds for arbitrary real-valued $h$, it follows that $\grad \NLTikhFunc$ is given by the expression in \eqref{eq:NLTikhFuncGradient}.

\section*{Funding}
The work was supported by the Deutsche Forschungsgemeinschaft (DFG, German Research
Foundation) through grant SFB 1456-C03.

\section*{Acknowledgements}
The authors thank Johannes Hagemann, Mareike T\"opperwien and Aike Ruhlandt for previous joint work, resulting in 
the data of Fig.1 to Fig.3, respectively, and for related fruitful discussions.
L.M.L.\ and T.S.\ are members of the Max Planck School of Photonics supported by BMBF, Max Planck Society and Fraunhofer Society.

\section*{Disclosures}
The authors declare no conflicts of interest.

\section*{Data Availability Statement}
Data underlying the results presented in this paper are not publicly available at this time but may be obtained from the authors upon reasonable request.

\bibliography{NLTikh}
\end{document}